\documentclass{moriond}

\def\be{\begin{equation}}
\def\ee{\end{equation}}
\def\bea{\begin{eqnarray}}
\def\eea{\end{eqnarray}}

\newcommand{\Photo}

\begin{document}

\definecolor{brown}{RGB}{139,69,19}

\newcommand{\nuc}[2]{$^{#2}\rm #1$}

\newcommand{\bb}[1]{$\rm #1\nu \beta \beta$}
\newcommand{\bbm}[1]{$\rm #1\nu \beta^- \beta^-$}
\newcommand{\bbp}[1]{$\rm #1\nu \beta^+ \beta^+$}
\newcommand{\bbe}[1]{$\rm #1\nu \rm ECEC$}
\newcommand{\bbep}[1]{$\rm #1\nu \rm EC \beta^+$}

\newcommand{\largeGERDA}{{LArGe}}
\newcommand{\PI}{\mbox{\textsc{Phase\,I}}}
\newcommand{\PIa}{\mbox{\textsc{Phase\,Ia}}}
\newcommand{\PIb}{\mbox{\textsc{Phase\,Ib}}}
\newcommand{\PIc}{\mbox{\textsc{Phase\,Ic}}}
\newcommand{\PII}{\mbox{\textsc{Phase\,II}}}


\newcommand{\fp}{F$_{\rm prompt}$}

\newcommand{\order}[1]{\mbox{$\mathcal{O}$(#1)}}

\newcommand{\pic}[5]{
       \begin{figure}[ht]
       \begin{center}
       \includegraphics[width=#2\textwidth, keepaspectratio, #3]{#1}
       \end{center}
       \caption{#5}
       \label{#4}
       \end{figure}
}

\newcommand{\apic}[5]{
       \begin{figure}[H]
       \begin{center}
       \includegraphics[width=#2\textwidth, keepaspectratio, #3]{#1}
       \end{center}
       \caption{#5}
       \label{#4}
       \end{figure}
}

\newcommand{\sapic}[5]{
       \begin{figure}[P]
       \begin{center}
       \includegraphics[width=#2\textwidth, keepaspectratio, #3]{#1}
       \end{center}
       \caption{#5}
       \label{#4}
       \end{figure}
}

\newcommand{\picwrap}[9]{
       \begin{wrapfigure}{#5}{#6}
       \vspace{#7}
       \begin{center}
       \includegraphics[width=#2\textwidth, keepaspectratio, #3]{#1}
       \end{center}
       \caption{#9}
       \label{#4}
       \vspace{#8}
       \end{wrapfigure}
}

\newcommand{\baseT}[2]{\mbox{$#1\times10^{#2}$}}
\newcommand{\baseTsolo}[1]{$10^{#1}$}
\newcommand{\THL}{$T_{\nicefrac{1}{2}}$}

\newcommand{\UBI}{$\rm cts/(kg \times yr \times keV)$}

\newcommand{\Uflux}{$\rm m^{-2} s^{-1}$}
\newcommand{\Ucpd}{$\rm cts/(kg \times d)$}
\newcommand{\Uexpo}{$\rm kg\times yr$}

\newcommand{\Qbb}{$Q_{\beta\beta}$}

\newcommand{\validate}{\textcolor{blue}{\textit{(validate!!!)}}}

\newcommand{\improve}{\textcolor{blue}{\textit{(improve!!!)}}}

\newcommand{\missing}{\textcolor{red}{\textbf{...!!!...} }}

\newcommand{\quanta}{\textcolor{red}{\textit{(quantitativ?) }}}

\newcommand{\misscite}{\textcolor{red}{[citation!!!]}}

\newcommand{\missref}{\textcolor{red}{[reference!!!]}\ }

\newcommand{\PC}{$N_{\rm peak}$}
\newcommand{\BIC}{$N_{\rm BI}$}
\newcommand{\PAPR}{$R_{\rm p/>p}$}

\newcommand{\PCR}{$R_{\rm peak}$}


\newcommand{\gline}{$\gamma$-line}
\newcommand{\glines}{$\gamma$-lines}

\newcommand{\gray}{$\gamma$-ray}
\newcommand{\grays}{$\gamma$-rays}

\newcommand{\bray}{$\beta$-ray}
\newcommand{\brays}{$\beta$-rays}

\newcommand{\betas}{$\beta$'s}


\newcommand{\tab}{\textcolor{brown}{Tab.~}}
\newcommand{\eq}{\textcolor{brown}{Eq.~}}
\newcommand{\fig}{\textcolor{brown}{Fig.~}}
\renewcommand{\sec}{\textcolor{brown}{Sec.~}}
\newcommand{\chap}{\textcolor{brown}{Chap.~}}

 \newcommand{\fn}{\iffalse \fi} 
 \newcommand{\tx}{\iffalse \fi} 
 \newcommand{\txe}{\iffalse \fi} 
 \newcommand{\sr}{\iffalse \fi} 

\vspace*{4cm}
\title{DEAP-3600 Recent Dark Matter Results}

\author{ B. Lehnert for the DEAP Collaboration}

\address{Carleton University, Department of Physics, \\1125 Colonel By Drive, Ottawa, (ON) K1S 5B6, Canada
}

\maketitle\abstracts{
The DEAP-3600 experiment is searching for WIMP dark matter with a 3.3 tonne single phase liquid argon (LAr) target, located at SNOLAB. The construction and filling of DEAP-3600 was completed in 2016, and the experiment is currently taking physics data. First results were recently published, which demonstrated stable detector operations and the power of pulse shape discrimination to reject electron-recoil backgrounds in LAr. In addition, the most sensitive WIMP exclusion with a LAr target was achieved at the time of publication. (Proceedings for 53$^{\rm rd}$ Rencontres de Moriond, Cosmology)}

\section{Introduction}

There is strong evidence that dark matter accounts for 26.8\% of the total energy budget of the universe and for 84.5\% of the total matter content \cite{Plank}. Weakly interacting massive particles (WIMPs) are among the favorite dark matter candidates which can be searched for with direct detection low background experiments. DEAP-3600 is a direct detection WIMP dark matter experiment based on a 3.3~tonne liquid argon (LAr) target. The detector setup is placed at SNOLAB in Sudbury, Canada, 2100~m below the earth surface.

WIMP dark matter particles from the galactic halo are expected to pass through the LAr target volume and recoil on \nuc{Ar}{40} nuclei, producing an exponential energy spectrum at the 10~keV$_{ee}$ scale in the detector. In oder to constrain the WIMP-nucleon cross section, the standard thermal galactic dark matter halo model with 544~km/s galactic escape velocity is assumed in addition to a dark matter density of 0.3~GeV/cm$^3$, along with the sun and earth velocities of 220~km/s and 230~km/s, respectively \cite{DMHaloModel}.

DEAP-3600 finished construction in 2016 and during the initial filling with LAr, a 4.4~d live-time dataset was recorded to demonstrate detector operations and obtain a first WIMP dark matter exclusion limit. In these proceedings, this dataset, the detector performance and analysis are presented. Ultimately, DEAP-3600 is designed to achieve a background level of less than 1 event in 3000~kg$\times$yr fiducial exposure in the region of interest (ROI) between 120 and 240 observed photoelectrons (PE). With this background-free exposure, a WIMP-nucleon cross section exclusion sensitivity of 10$^{-46}$ cm$^2$ is expected at 100 GeV/c$^2$ WIMP mass.

\section{DEAP-3600 Detector and Background Mitigation}

The detector is described in detail in \cite{DEAPDetector}. The target material is contained in a spherical acrylic vessel (AV) of 85~cm radius able to hold 3600~kg of LAr. For the commissioning dataset, the AV was filled with $3322\pm110$~kg of LAr. The inner surface of the AV was removed in order to mitigate Rn daughters diffused into the outer layers during construction. The surface was then coated with a TPB layer which serves as a wavelength shifter for the 128~nm LAr scintillation light. The shifted light is observed by 255 Hamamatsu R5912 HQE PMTs concentrically arranged around the AV, 50~cm apart. The light passes through acrylic light guides optically connecting the PMTs with the AV. They serve as a passive neutron shield as well as to keep a thermal gradient between LAr (87~K) and PMTs (273 K). The whole setup is enclosed inside a stainless steel shell and immersed in a cylindrical 7.8~m diameter and 7.8~m high water tank. The water serves as a passive shield for outside \grays\ and neutrons and is instrumented with 48 outward looking PMTs as a muon Cherenkov veto. While filling, the LAr passes through a process system including a SAES getter and Rn trap to remove chemical and radioactive impurities. It enters the detector through a 30~cm wide neck on top of the AV and is hermetically sealed inside during normal operations.

The main background reduction is based on the powerful pulse shape discrimination (PSD) in LAr scintillation. Short singlet (6~ns) and long triplet (1300~ns) dimer excitations are populated differently for background-like electronic recoil (ER) events and WIMP-like nuclear recoil (NR) events. A simple PSD parameter \fp\ is defined as the ratio of observed PE in the first 150~ns w.r.t.\ the light in 10~$\mu$s of the event. ER rejection efficiencies up to \baseTsolo{10} can be reached. In atmospheric LAr, such a high rejection power is needed in order to mitigate the high activity of the cosmogenic isotope \nuc{Ar}{39} at about 1~Bq/kg of LAr. In DEAP-3600 about 0.2 \nuc{Ar}{39} background events for the WIMP search are expected in the design exposure of 3000~kg$\times$yr at a 50\% NR acceptance. Other main background sources are neutron interactions, which create NRs similar to WIMPs, and alpha emitters in the surface layers of the detector for which only a fraction of the alpha energy generates scintillation light. Both contributions are expected to be below 0.2 events in the WIMP search region within the design exposure.

\section{Commissioning Data}

\begin{figure}
\centerline{\includegraphics[width=0.8\linewidth]{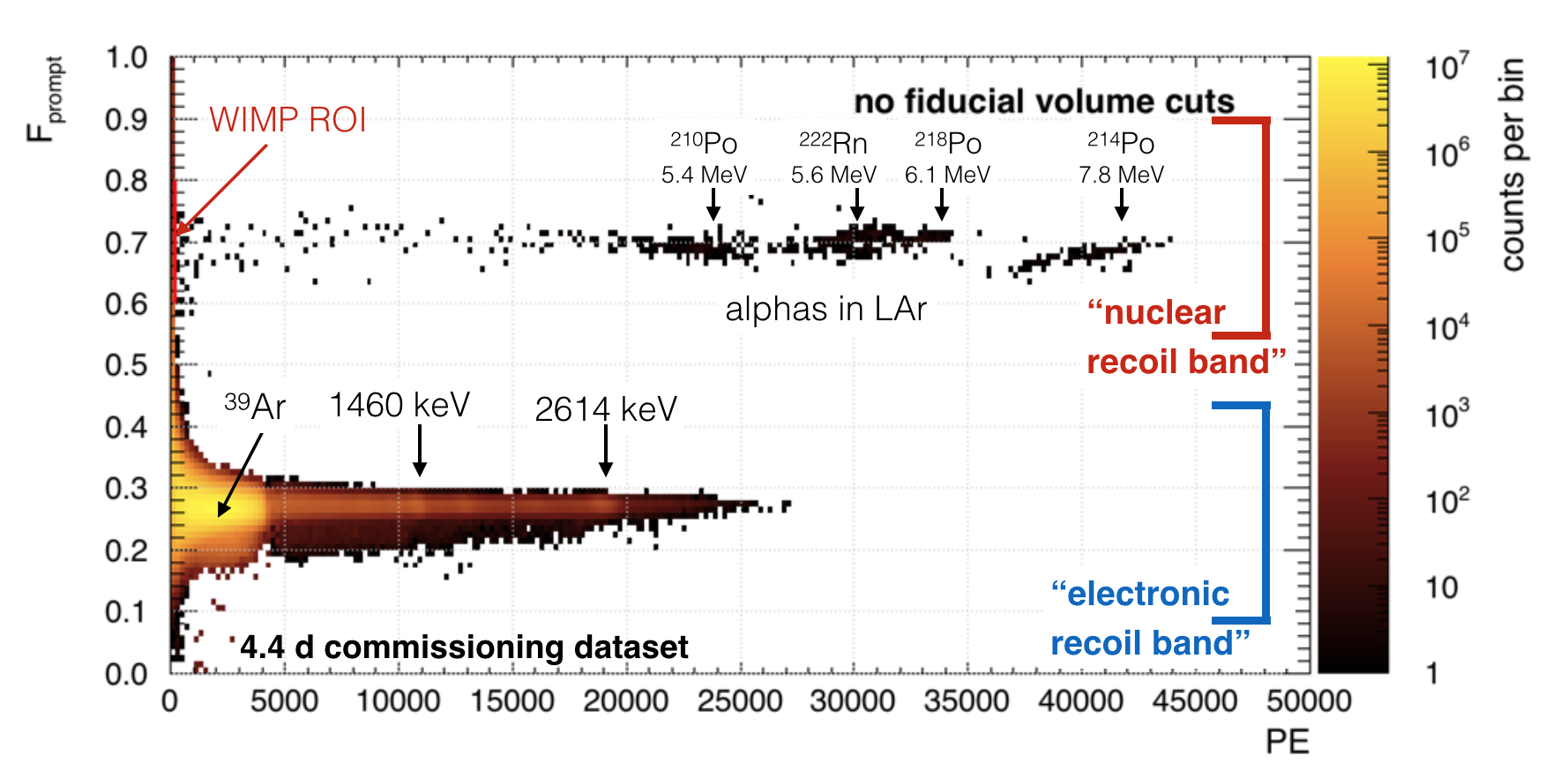}}
\caption{ PSD (\fp) vs energy (PE) plane of the 4.4~d dataset before fiducializing cuts.}
\label{pic_fpVsPE}
\end{figure}

A 10~d stable period during the filling of the detector in June 2016 was used for analysis. 4.7~d of runtime were selected with stability criteria resulting in 4.4~d live time, or 9.9~tonne$\times$d fiducial exposure \cite{DEAPPRL}. 
\fig \ref{pic_fpVsPE} shows the \fp\ versus detected PE for the dataset. Two bands emerge: nuclear recoils at high \fp\ and electronic recoils at low \fp. Prominent background features are highlighted. NRs at high energies are alpha emitters in the bulk LAr from \nuc{Rn}{222} and its daughters. A specific activity of $(1.8\pm0.2)\times10^{-1}$~$\mu$Bq/kg is observed for \nuc{Rn}{222}. A peak of \nuc{Po}{210} is observed at the surface of the detector, which is fed by residual long-lived \nuc{Pb}{210} implanted into the detector surface during construction. From construction history, the origin is assumed to be between the TPB and acrylic surfaces and an activity of $0.22\pm0.04$~mBq/m$^2$ is determined for this hypothesis. Furthermore, a contribution from the acrylic bulk could be constrained to be less than 3.3~mBq in the first 80~$\mu$m. The ER band is dominated by \nuc{Ar}{39} below 565~keV roughly equivalent to 4500 PE. Above that energy \glines\ from natural decay chains and \nuc{K}{40} are observed as well as betas emitted from \nuc{K}{42} \cite{BGTAUPProceeding}.

\begin{figure}
\centerline{\includegraphics[width=0.99\linewidth]{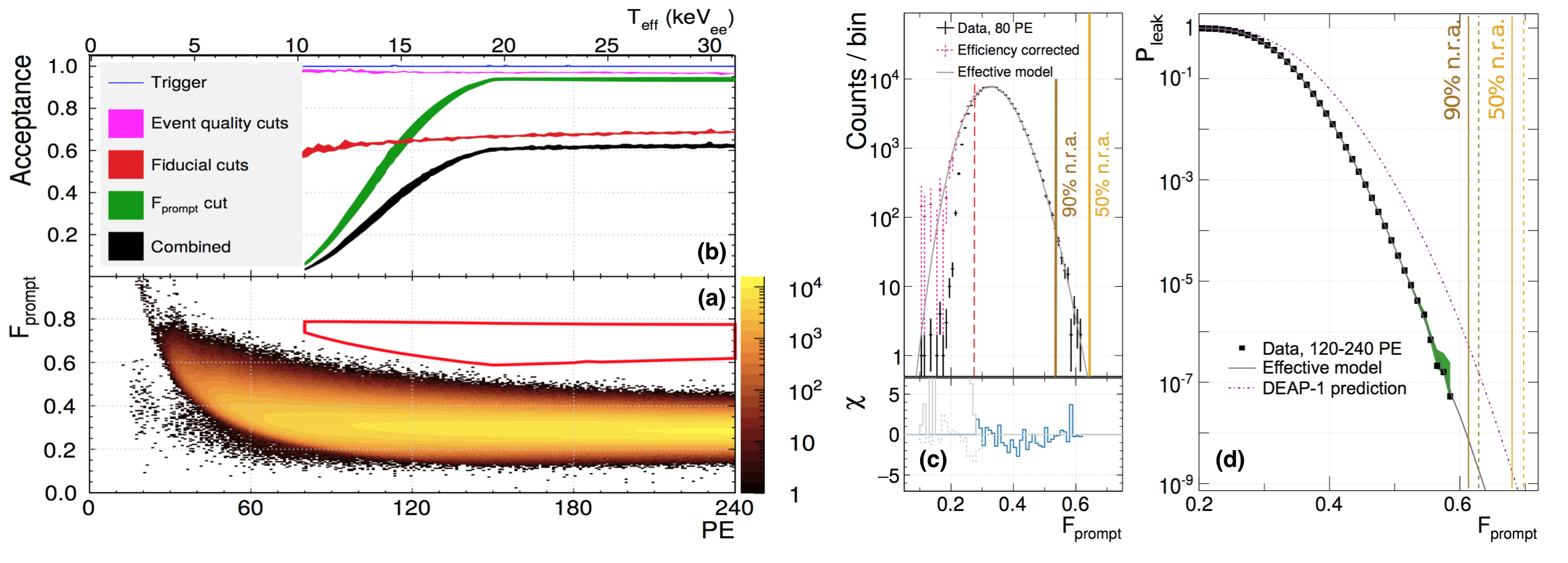}}
\caption{Region of interest for dark matter search $^3$. (a) definition of ROI and surviving event in the \fp\ vs PE plane after all cuts. (b) energy dependence of cut acceptances. (c) 1~PE \fp\ slice at 80~PE including PSD fit model. (d) ER leakage fraction in 120 - 240~PE including PSD fit model and predictions based on R\&D.}
\label{pic_ROIComposite}
\end{figure}

The ROI for the WIMP search is shown in \fig \ref{pic_ROIComposite} (a) as the red area between 80 and 240~PE. The low \fp\ bound is chosen such that a total of 0.2 \nuc{Ar}{39} background events are expected in the ROI predicted by the PSD model \cite{DEAP1} or that a WIMP acceptance of 95\% is reached for a given PE bin. The PSD model is validated down to 80~PE, which defines the low energy bound. The high energy bound at 240~PE is chosen ad-hoc in order to minimize potential backgrounds in a region with low WIMP sensitivity. The high \fp\ bound removes potential background events due to pile-up with Cherenkov light, while keeping 99\% of the WIMP acceptance. The energy dependence of the cut acceptances are shown in \fig \ref{pic_ROIComposite} (b). A total flat NR acceptance of 66.9\% is determined for all but the \fp\ cut. 

The PSD model is illustrated in \fig \ref{pic_ROIComposite} (c) for a 1~PE \fp\ slice at 80~PE. The PSD model is fit to data above the vertical dashed line. Below this line the \fp\ distribution is influenced by a trigger efficiency $<1$, as illustrated. The brown and yellow vertical lines correspond to the 90\% and 50\% NR acceptance. \fig \ref{pic_ROIComposite} (d) shows the leakage fraction of ER events between 120 and 240 PE for this dataset as well as for a prediction from the R\&D experiment DEAP-1 \cite{DEAP1}. The observed PSD leakage is smaller than  predicted due to a lower than expected electronic noise in DEAP-3600. A worldwide best leakage of  \baseT{<1.2}{-7} (90\% C.L.) at 90\% NR acceptance is demonstrated.

The energy scale and resolution were calibrated with an \nuc{Ar}{39} spectral fit to a 13~min subset of the commissioning data. This fit was combined with a \nuc{Na}{22} calibration run after detector modifications. Differences due to these modifications were taken into account as systematic uncertainties. 
At 80~keV a PE yield of $7.80\pm0.43$ PE/keV$_{ee}$ is determined and a resolution of $20\pm1$\% is observed in the fits. However, due to limited sensitivity to the resolution in the fit to the continuous spectra, a conservative lower resolution bound of 12\% based on counting statistics and single PE calibrations is used in the analysis. Due to the steeply falling exponential WIMP recoil spectra, a wider resolution would increase the sensitivity. The energy scale is illustrated in the left plot of \fig \ref{pic_ROIComposite}.

\begin{figure}
\centerline{\includegraphics[width=0.6\linewidth]{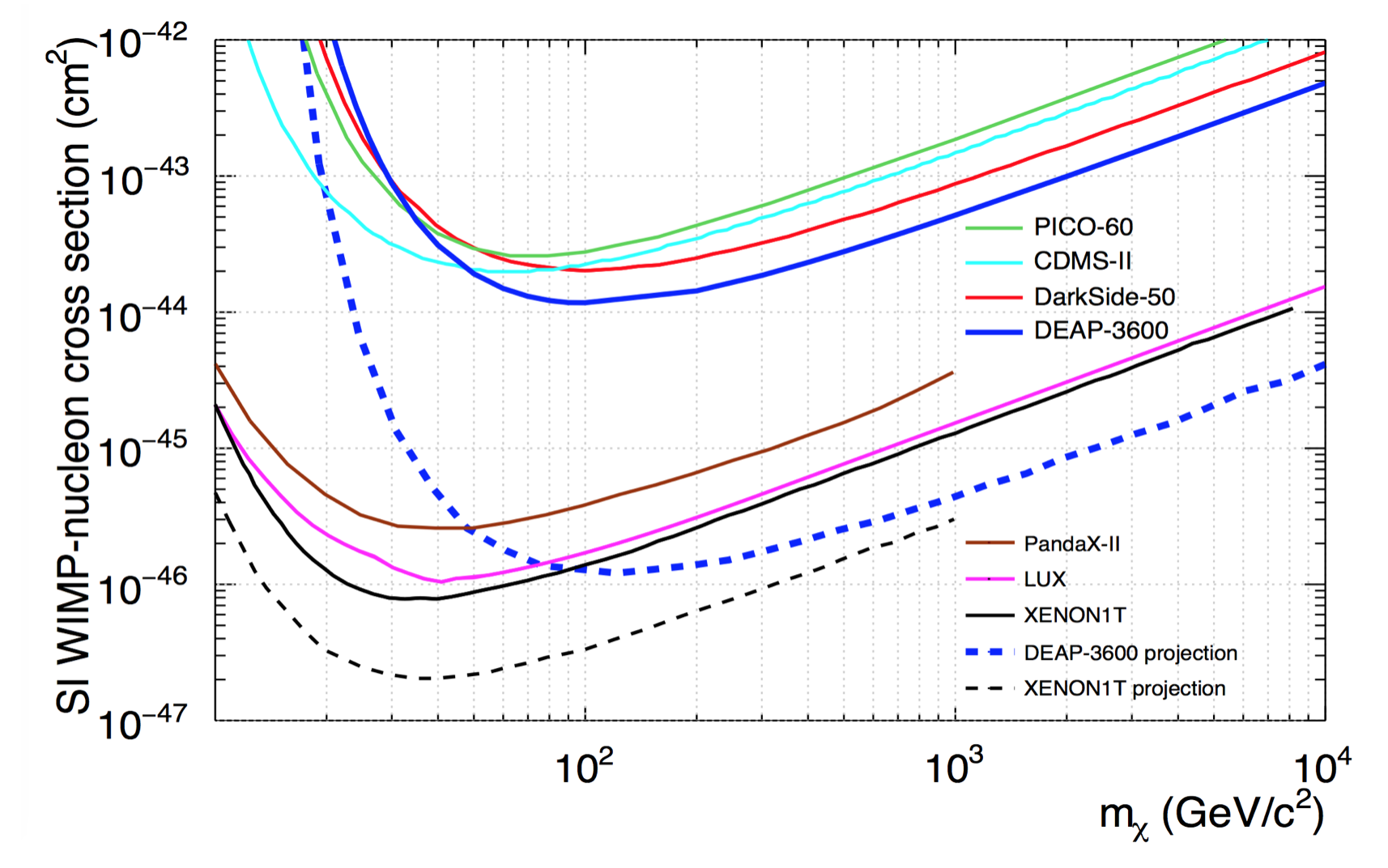}}
\caption{WIMP exclusion curve obtained in this work (solid blue) and by other experiments (solid lines). The predictions for DEAP-3600 and Xenon1T are also shown (dashed lines). References in $^3$.}
\label{pic_WIMPExclusion}
\end{figure}

Zero events are observed in the ROI after all cuts. The largest expected background are 0.2~events from \nuc{Ar}{39}. With a conservative assumption of zero observed and zero expected events, a WIMP-nucleon interaction cross section of \baseT{1.2}{-44}~ cm$^2$ at 100 GeV/c$^2$ could be excluded at 90\% CL. The limit includes systematic uncertainties from the total LAr mass, the \fp\ cut acceptance, PSD independent cut acceptances, the energy scale and the energy resolution. 
The exclusion curve from this work, from other leading WIMP dark matter experiments as well as projected sensitivities are shown in \fig \ref{pic_WIMPExclusion}.  

\section{Conclusions and Outlook}

DEAP-3600 has demonstrated the successful and stable operation of a single phase tonne-scale LAr WIMP dark matter detector. 
After this dataset was taken, a contamination of the LAr occurred and the detector was refilled in Oct.\ 2016. Since Nov.\ 2016, the detector has operated stably with  $3256\pm110$~kg LAr, and a 1~yr dataset was recorded without a blinding scheme. This dataset is currently being analyzed. Since Jan.\ 2018 a blinding scheme has been applied. 
The initial commissioning dataset could demonstrate better than expected PSD performance as well as the lowest \nuc{Rn}{222} concentration in a large noble liquid dark matter experiment to date \cite{DEAPPRL}. The achieved WIMP exclusion limit with only 4.4~d was the best to date for a LAr target and was only recently surpassed by Darkside-50 with a 532~d dataset \cite{DS50}.

Recently, a worldwide LAr dark matter community was formed driving the development of direct WIMP searches with LAr target toward the natural detection limit at the neutrino floor. First, DarkSide 20k, a two phase TPC with about 20~tonnes underground LAr is planned at LNGS \cite{DS20k}, followed by an envisioned but yet unspecified 300~tonne LAr detector.

\end{document}